\documentclass[reprint,amsmath,prl]{revtex4-1} 

\usepackage{color}
\definecolor{red}{rgb}{1,0,0}
\definecolor{blue}{rgb}{0,0,1}

\usepackage[utf8]{inputenc}
\usepackage{siunitx} 
\usepackage[]{natbib}
\usepackage{amssymb}
\usepackage{mathtools} 
\usepackage{booktabs} 
\usepackage{xspace} 
\usepackage{txfonts}
\usepackage{times}
\DeclareMathAlphabet{\mathpzc}{OT1}{pzc}{m}{it}

\usepackage{graphicx}
\usepackage[]{subcaption} 
\ifx\DOLINENUMBERS

\newcommand*\patchAmsMathEnvironmentForLineno[1]{%
  \expandafter\let\csname old#1\expandafter\endcsname\csname #1\endcsname
  \expandafter\let\csname oldend#1\expandafter\endcsname\csname end#1\endcsname
  \renewenvironment{#1}%
     {\linenomath\csname old#1\endcsname}%
     {\csname oldend#1\endcsname\endlinenomath}}%
\newcommand*\patchBothAmsMathEnvironmentsForLineno[1]{%
  \patchAmsMathEnvironmentForLineno{#1}%
  \patchAmsMathEnvironmentForLineno{#1*}}%
\AtBeginDocument{%
\patchBothAmsMathEnvironmentsForLineno{equation}%
\patchBothAmsMathEnvironmentsForLineno{align}%
\patchBothAmsMathEnvironmentsForLineno{flalign}%
\patchBothAmsMathEnvironmentsForLineno{alignat}%
\patchBothAmsMathEnvironmentsForLineno{gather}%
\patchBothAmsMathEnvironmentsForLineno{multline}%
}

\fi

\def\pqset{\lbrace p,q \rbrace\xspace}

\renewcommand{\Re}{\operatorname{Re}}

\usepackage{ifthen} 

\newcommand{\eqnref}[1]{(\ref{#1})} 
\gdef\vect#1{\mathbf{#1}} 

\gdef\genexpEsymb{\alpha}
\gdef\genexpHsymb{\beta}
\gdef\pseudosymb{(\genexpEsymb)}

\newcommand{\genE}[1]{\lambda^{#1} } 
\newcommand{\genH}[1]{\lambda^{#1} } 

\gdef\Einvarreal{E}
\gdef\Einvarpseudo{E^{\pseudosymb}}

\gdef\Hinvarreal{H}

\newcommand{\MYVAR}[5]{ 
	\ifthenelse{\equal{#2}{}}{\vect{#1}}{#1_{#2}}^{#3}
	\ifthenelse{\equal{#4}{}}{}{(\ifthenelse{\equal{#5}{}}{\vect{#4}}{#4_{#5}})} 
}
\newcommand{\hs}[3]    {\vect{h}_{#1}^{#2}\ifthenelse{\equal{#3}{}}{}{(\vect{#3})}} 
 
\newcommand{\uu}[4]{\MYVAR{u}{#1}{#2}{#3}{#4}} 
\newcommand{\uf}[2]{\uu{#1}{}{#2}{}}  
\newcommand{\sprime}{^{\prime}\mkern-1.2mu}
\newcommand{\dprime}{^{\prime\prime}\mkern-1.2mu}

\newcommand{\ush}[3]{ \ifthenelse{\equal{#3}{1}}{u_{#1}^{#2,*} }{ u_{#1}^{#2} } }

\gdef\Sk{s}
\gdef\Sp{s\sprime}
\gdef\Sq{s\dprime}

\gdef\VK{\vect{k}}
\gdef\VP{\vect{k\sprime}}
\gdef\VQ{\vect{k\dprime}}

\gdef\VKLEN{k}
\gdef\VPLEN{k\sprime}
\gdef\VQLEN{k\dprime}

\gdef\KIDX{n}
\newcommand{\krotidx}[1]{
	\ifthenelse{\equal{#1}{1}}{\KIDX}{
	\ifthenelse{\equal{#1}{2}}{\KIDX}{
	\ifthenelse{\equal{#1}{3}}{\KIDX}{
	\ifthenelse{\equal{#1}{-1}}{\KIDX-2}{
	\ifthenelse{\equal{#1}{-2}}{\KIDX-1}{
	\ifthenelse{\equal{#1}{-3}}{\KIDX}{
	}}}}}}
}
\newcommand{\protidx}[1]{
	\ifthenelse{\equal{#1}{1}}{\KIDX+p}{
	\ifthenelse{\equal{#1}{2}}{\KIDX-p}{
	\ifthenelse{\equal{#1}{3}}{\KIDX-q}{
	\ifthenelse{\equal{#1}{-1}}{\KIDX-1}{
	\ifthenelse{\equal{#1}{-2}}{\KIDX-2}{
	\ifthenelse{\equal{#1}{-3}}{\KIDX-2}{
	}}}}}}
}
\newcommand{\qrotidx}[1]{
	\ifthenelse{\equal{#1}{1}}{\KIDX+q}{
	\ifthenelse{\equal{#1}{2}}{\KIDX+q-p}{
	\ifthenelse{\equal{#1}{3}}{\KIDX+p-q}{
	\ifthenelse{\equal{#1}{-1}}{\KIDX}{
	\ifthenelse{\equal{#1}{-2}}{\KIDX}{
	\ifthenelse{\equal{#1}{-3}}{\KIDX-1}{
	}}}}}}
}

\newcommand{\krotidxshifted}[1]{
	\ifthenelse{\equal{#1}{1}}{\KIDX-q}{
	\ifthenelse{\equal{#1}{2}}{\KIDX-q+p}{
	\ifthenelse{\equal{#1}{3}}{\krotidx{3}}{
	}}}
}
\newcommand{\protidxshifted}[1]{
	\ifthenelse{\equal{#1}{1}}{\KIDX-q+p}{
	\ifthenelse{\equal{#1}{2}}{\KIDX-q}{
	\ifthenelse{\equal{#1}{3}}{\protidx{3}}{
	}}}
}
\newcommand{\qrotidxshifted}[1]{
	\ifthenelse{\equal{#1}{1}}{\KIDX}{
	\ifthenelse{\equal{#1}{2}}{\KIDX}{
	\ifthenelse{\equal{#1}{3}}{\qrotidx{3}}{
	}}}
}

\gdef\eqsgn#1#2{   \ifthenelse{\equal{#1}{#2}}{+}{-}}
\gdef\eqsgnnp#1#2{ \ifthenelse{\equal{#1}{#2}}{ }{-}}
\gdef\eqsgninv#1#2{\ifthenelse{\equal{#1}{#2}}{-}{+}}

\newcommand{\effsign}[2]{
	\StrLeft{#1}{1}[\firstletterA]
	\StrLeft{#2}{1}[\firstletterB]
	\ifthenelse{\equal{\firstletterA}{\firstletterB}}{+}{-}
}

\newcommand{\effsymnp}[2]{
	\StrLeft{#1}{1}[\firstletterA]
	\StrLeft{#2}{1}[\firstletterB]
	\StrLen{#2}[\lengthB]
	\ifthenelse{\equal{\firstletterA}{\firstletterB}}{
		\ifthenelse{\equal{\lengthB}{1}}{+}{} 
	}{-} 
	\StrGobbleLeft{#2}{1}
}

\gdef\Wepsfunc#1#2{\epsilon_{#1}^{#2}}

\gdef\Wepssubmodel{\Wepsfunc{p,q}{}}

\gdef\Wxifunc#1#2{\xi_{#1}^{#2}}

\gdef\WxideffuncforEgeneqn#1#2#3{(#2\epsilon_{#1}^{#2,#3}-1)}

\gdef\Egeneqn#1#2#3{1 - \Sp\ifthenelse{\equal{#1}{1}}{\genE{#3}}{(\genE{#3})^{#1}}
\ifthenelse{\equal{#1}{1}}{\Wepsfunc{}{\Sp,\Sq}}{\Wepsfunc{#1,#2}{\Sp,\Sq}}
+ 
(\genE{#3})^{#2}
\ifthenelse{\equal{#1}{1}}{\Wxifunc{}{\Sp,\Sq}}{\WxideffuncforEgeneqn{#1,#2}{\Sp}{\Sq}}
}

\gdef\Hgeneqn#1#2#3{1 -    \ifthenelse{\equal{#1}{1}}{\genH{#3}}{(\genH{#3})^{#1}}
\ifthenelse{\equal{#1}{1}}{\Wepsfunc{}{\Sp,\Sq}}{\Wepsfunc{#1,#2}{\Sp,\Sq}}
+    (\genH{#3})^{#2}
\ifthenelse{\equal{#1}{1}}{\Wxifunc{}{\Sp,\Sq}}{\Wxifunc{#1,#2}{\Sp,\Sq}}
}

\gdef\helampslong#1#2#3#4#5#6{u_{#1}^{#2}(#3)u_{#4}^{#5}(#6)} 
\newcommand{\helampsshort}[6]{u_{#3}^{#1#2}u_{#6}^{#4#5}} 
\newcommand{\hamps}[5]{
	\ifthenelse{\equal{#1}{0}}{ \helampslong{#2}{ }{#3}{#4}{ }{#5} }{ 
	\ifthenelse{\equal{#1}{1}}{ \helampslong{#2}{*}{#3}{#4}{ }{#5} }{ 
	\ifthenelse{\equal{#1}{2}}{ \helampslong{#2}{*}{#3}{#4}{ }{#5} }{ 
	\ifthenelse{\equal{#1}{3}}{ \helampslong{#2}{ }{#3}{#4}{ }{#5} }{ 
	\ifthenelse{\equal{#1}{-1}}{ \helampsshort{#2}{,*}{#3}{#4}{}{#5} }{ 
	\ifthenelse{\equal{#1}{-2}}{ \helampsshort{#2}{,*}{#3}{#4}{}{#5} }{ 
	\ifthenelse{\equal{#1}{-3}}{ \helampsshort{#2}{}{#3}{#4}{}{#5} }{ 
	}}}}}}}
}

\gdef\KIDXs{n}
\newcommand{\tripleintA}[6]{
\ush{\ifthenelse{\equal{#1}{}}{\KIDXs}{\KIDXs#1}}{#4}{1}    
\ush{\ifthenelse{\equal{#1}{}}{\KIDXs+1}{\KIDXs#2}}{#5}{1} 
\ush{\ifthenelse{\equal{#1}{}}{\KIDXs+2}{\KIDXs#3}}{#6}{0}}
\newcommand{\tripleintB}[6]{
\ush{\ifthenelse{\equal{#1}{}}{\KIDXs-1}{\KIDXs#1}}{#4}{1}  
\ush{\ifthenelse{\equal{#1}{}}{\KIDXs}{\KIDXs#2}}{#5}{1}   
\ush{\ifthenelse{\equal{#1}{}}{\KIDXs+1}{\KIDXs#3}}{#6}{0}}
\newcommand{\tripleintC}[6]{
\ush{\ifthenelse{\equal{#1}{}}{\KIDXs-2}{\KIDXs#1}}{#4}{0}  
\ush{\ifthenelse{\equal{#1}{}}{\KIDXs-1}{\KIDXs#2}}{#5}{0} 
\ush{\ifthenelse{\equal{#1}{}}{\KIDXs}{\KIDXs#3}}{#6}{1}}

\newcommand{\tripleintAconj}[6]{
\ush{\ifthenelse{\equal{#1}{}}{\KIDXs}{\KIDXs#1}}{#4}{0}    
\ush{\ifthenelse{\equal{#1}{}}{\KIDXs+1}{\KIDXs#2}}{#5}{0} 
\ush{\ifthenelse{\equal{#1}{}}{\KIDXs+2}{\KIDXs#3}}{#6}{1}}
\newcommand{\tripleintCb}[6]{
\ush{\ifthenelse{\equal{#1}{}}{\KIDXs-2}{\KIDXs#1}}{#4}{1} 
\ush{\ifthenelse{\equal{#1}{}}{\KIDXs-1}{\KIDXs#2}}{#5}{1} 
\ush{\ifthenelse{\equal{#1}{}}{\KIDXs}{\KIDXs#3}}{#6}{0}}

\newcommand{\modalcontr}[1]{(\vert\ush{\KIDXs}{+}{0}\vert^2 #1 \vert\ush{\KIDXs}{-}{0}\vert^2)}
\newcommand{\triplehelampsshort}[9]{
u_{#1}^{#3\ifthenelse{\equal{#2}{}}{}{,#2}}
u_{#4}^{#6\ifthenelse{\equal{#5}{}}{}{,#5}}
u_{#7}^{#9\ifthenelse{\equal{#8}{}}{}{,#8}}
}

\gdef\meanenergy{\epsilon_{\mathrm{in}}}

\gdef\sigmastate{\lbrace\Sk,\Sp,\Sq\rbrace}
\gdef\sigmastateA{\pm\lbrace+,-,+\rbrace}
\gdef\sigmastateB{\pm\lbrace+,-,-\rbrace}
\gdef\sigmastateC{\pm\lbrace+,+,-\rbrace}
\gdef\sigmastateD{\pm\lbrace+,+,+\rbrace}

\begin{document}

\title{Pseudo-invariants causing inverse energy cascades in three-dimensional turbulence}

\author{Nicholas M. Rathmann}
\email{rathmann@nbi.ku.dk}
\affiliation{Niels Bohr Institute, University of Copenhagen, Denmark}
\author{Peter D. Ditlevsen}
\email{pditlev@nbi.ku.dk}
\affiliation{Niels Bohr Institute, University of Copenhagen, Denmark}
	
\begin{abstract}
Three-dimensional (3D) turbulence is characterized by a dual forward cascade of both kinetic energy and helicity, a second inviscid flow invariant, from the integral scale of motion to the viscous dissipative scale. In helical flows, however, such as strongly rotating flows with broken mirror symmetry, an inverse energy cascade can be observed analogous to that of two-dimensional turbulence (2D) where a second positive-definite flow invariant, enstrophy, unlike helicity in 3D, effectively blocks the forward cascade of energy. In the spectral--helical decomposition of the Navier--Stokes equation it has previously been show that a subset of three-wave (triad) interactions conserve helicity in 3D in a fashion similar to enstrophy in 2D, thus leading to a 2D-like inverse energy cascade in 3D.
In this work, we show both theoretically and numerically that an additional subset of interactions exist conserving a new pseudo-invariant in addition to energy and helicity, which contributes either to a forward or inverse energy cascade depending on the specific triad interaction geometry.
\end{abstract}
\maketitle

Fully developed three-dimensional (3D) turbulence is characterised by a forward cascade of kinetic energy from the large integral scale of motion to the small Kolmogorov scale of viscous dissipation.  
In the large Reynolds number limit,
$\eta \rightarrow 0$, the production of enstrophy, the integral of the vorticity squared, by the stretching and bending term in the incompressible Navier--Stokes equations (NSE) permits
the viscous dissipation of energy at the Kolmogorov scale.
In two-dimensional (2D) turbulence, the stretching and bending term is absent, and enstrophy is, in addition to energy, also an inviscid 
invariant \citep{bib:kraichnan1967inertial}.
In this case, the dissipation of enstrophy prevents a simultaneous dissipation of energy at the Kolmogorov scale, effectively blocking the
forward cascade of energy. 
The dual inviscid conservation of both quantities, $\int E(k)dk$ and $\int k^2E(k)dk$, the integrals over the spectral energy and enstrophy densities, consequently implies a reversal of the energy cascade to larger scales (inverse cascade).
In 3D turbulence, helicity, the integral of the scalar product of velocity and vorticity, is also an inviscid invariant \citep{bib:moffatt1969degree}. 
Similarly to the enstrophy spectrum, the helicity spectrum $H(k)\sim kE(k)$ dominates over the energy spectrum at small scales (large $k$), but unlike enstrophy, helicity is not sign definite. 
As a consequence, the increased dissipation of both signs of helicity compared to energy can be obtained without a net helicity production as long as the dissipation of both positive and negative helicities balance \cite{bib:ditlevsen2001dissipation}. Inviscid conservation of helicity does therefore not prevent a forward cascade of energy \citep{bib:brissaud1973helicity}. 

In helical flows, however, such as strongly rotating flows with broken mirror symmetry, a simultaneous forward helicity cascade and inverse energy cascade can be observed \cite{bib:mininni2009scale}. 
In the spectral decomposition of the NSE, energy and helicity (and enstrophy in 2D) are conserved within each three-wave interaction (triad interaction).
It was recently proposed that inverse energy cascades might be a general characteristic of 3D turbulence \cite{bib:biferale2012inverse}, carried by a specific subset of triad interactions among helical wave components \citep{bib:waleffe1992nature} of the same sign which render helicity enstrophy-like. 
The relative roles played by the different of subsets of helical triad interactions would 
depend specifically on symmetries and boundary conditions of the turbulent flow \citep{bib:rathmannditlevsen2016,bib:alexakis2016helically}.  

\begin{figure*}[!t]
\begin{center}
\includegraphics[scale=1]{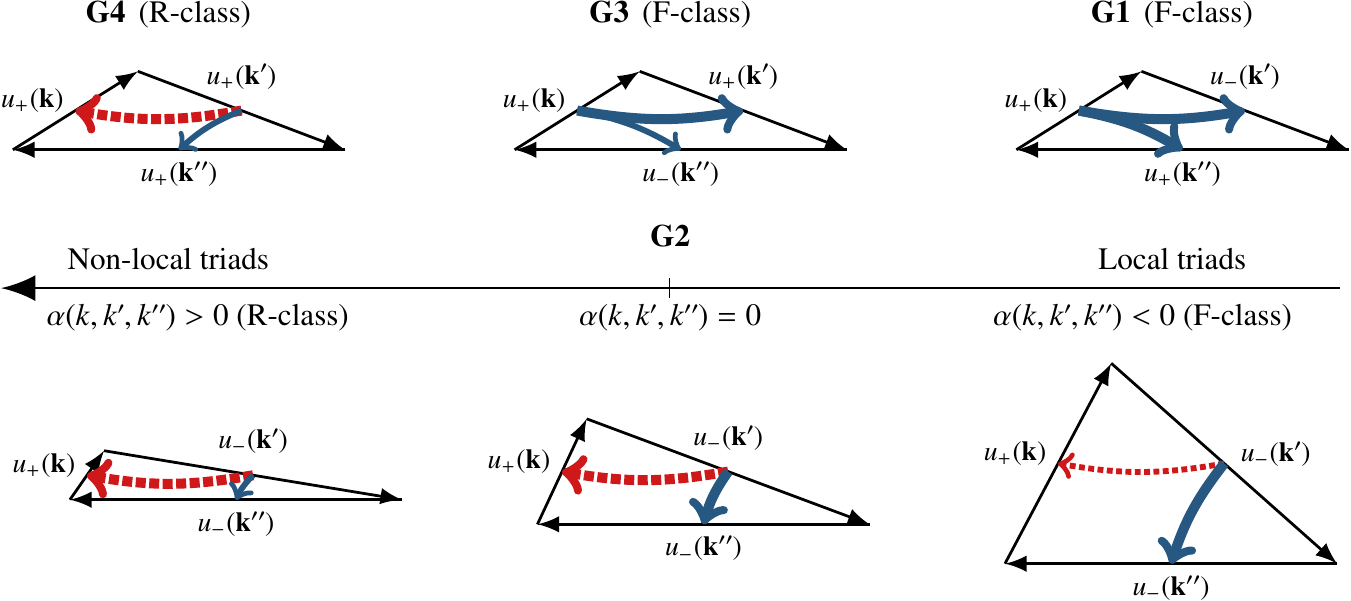}
\end{center}
\caption[helicalcouplings]{
G1--G4 helical interactions classified by behaviour (F- and R-classes). 
The behaviour of G2 is determined by the conservation of the new geometry-dependant enstrophy-like quantity $E^{(\alpha)}$.
The arrows indicate the average energy transfer directions based on a linear stability analysis \citep{bib:waleffe1992nature}: Blue/solid (red/dashed) arrows denote forward (reverse) energy transfers whilst thick (thin) arrows represent the dominant (subordinate) transfers.\label{fig:helicalcouplings}}
\end{figure*}
Applying the helical decomposition \cite{bib:waleffe1992nature} to the NSE, triad interactions are split into four distinct groups of interactions depending on the relative weights of interchange of energy and
helicity among the three waves.
Within each interaction group, we show that an additional either helicity- or enstrophy-like quantity is conserved. 
In this work, we conjecture that it is the spectral properties of this triad-specific invariant that governs the dual cascade of energy and helicity
in 3D turbulence. 
Our conjecture is confirmed in the case of a shell model which obeys the same conservations as the NSE \cite{bib:rathmannditlevsen2016}.  
\gdef\Edef{E=\sum_\vect{k}  (\vert u_{+}(\VK) \vert ^2 + \vert u_{-}(\VK) \vert^2)}
\gdef\Hdef{H=\sum_\vect{k} k(\vert u_{+}(\VK) \vert ^2 - \vert u_{-}(\VK) \vert^2)}

In the helical decomposition \cite{bib:waleffe1992nature} 
of the NSE for incompressible flows, each complex spectral velocity component $\vect{u}(\VK)$ is decomposed into helical helical modes $\vect{h}_\pm(\VK)$ (using $\VK\cdot\vect{u}(\VK) = 0$) which are complex eigenmodes of the curl operator, i.e. $i \VK \times \vect{h}_\pm = \pm k \vect{h}_\pm$, where $k=\vert\VK\vert$ is the length of the wave vector $\VK$.  In this basis, velocity components are given by $\vect{u}(\VK) = u_+(\VK)\vect{h}_+ + u_-(\VK)\vect{h}_-$, and energy and helicity are given by
$\Edef$ and $\Hdef$, respectively. The spectral NSE become  \citep{bib:waleffe1992nature}
\gdef\hhhweight{\hs{\Sp}{*}{\VP}\times\hs{\Sq}{*}{\VQ}\cdot\hs{\Sk}{*}{\VK}}
\begin{align}
(d_t + \nu k^2)\uf{\Sk}{k} = -1/4 \sum\limits_{\VK+\VP+\VQ=0}\;\sum_{\Sp,\Sq} (\Sp\VPLEN - \Sq \VQLEN)\notag\\ \hhhweight \; \uu{\Sp}{*}{\VPLEN}{} \uu{\Sq}{*}{\VQLEN}{}\label{eqn:Helical_NSE},
\end{align}
where $\lbrace \Sk,\Sp,\Sq \rbrace = \pm1$ are the helical signs of the interacting modes and $(\Sp\VPLEN - \Sq \VQLEN)\, \hhhweight$ is the coupling coefficient of the helical triad interaction involving velocity components $\lbrace\uu{\Sk}{}{\VKLEN}{},\uu{\Sp}{}{\VPLEN}{},\uu{\Sq}{}{\VQLEN}{}\rbrace$.
Each triad interaction in the spectral NSE is thus split into eight helical triad interactions by the inner sum over helical signs in \eqnref{eqn:Helical_NSE}. 
By sorting these interactions, four pairs with similar interaction coefficients arise:  
$\sigmastate = \sigmastateA,\sigmastateB,\sigmastateC,\sigmastateD$, hereafter referred to as groups G1-G4 respectively, see FIG. \ref{fig:helicalcouplings}. 

Isolating terms in \eqnref{eqn:Helical_NSE} involving only three wave vectors $\lbrace\VK,\VP,\VQ\rbrace$ (a single triad), and defining the shorthand notation $g=\hhhweight$, one finds, using the cyclic property of $g$,
\begin{align} 
d_t \uf{\Sk}{\VKLEN} &=(\Sp\VPLEN - \Sq\VQLEN)\,g\,\uu{\Sp}{*}{\VPLEN}{}\uu{\Sq}{*}{\VQLEN}{}\notag\\
d_t \uf{\Sp}{\VPLEN} &=(\Sq\VQLEN - \Sk\VKLEN)\,g\,\uu{\Sq}{*}{\VQLEN}{}\uu{\Sk}{*}{\VKLEN}{}\label{eqn:triadeqns}\\
d_t \uf{\Sq}{\VQLEN} &=(\Sk\VKLEN - \Sp\VPLEN)\,g\,\uu{\Sk}{*}{\VKLEN}{}\uu{\Sp}{*}{\VPLEN}{}.\notag
\end{align}
This simple form of the helically decomposed NSE triad dynamics is the basis of our analysis. 
Note that the cyclic symmetry of
\eqnref{eqn:triadeqns} implies that one may assume $k\le k'\le k''$ without loss of generality.
Multiplying by $\uu{\Sk}{*}{\VKLEN}{}, \uu{\Sp}{*}{\VPLEN}{}$ and $\uu{\Sq}{*}{\VQLEN}{}$, respectively, in the three equations \eqnref{eqn:triadeqns}, it immediately follows that energy is conserved within each triad interaction, and similarly for helicity by multiplication of $\Sk\VKLEN\uu{\Sk}{*}{\VKLEN}{}, \Sp\VPLEN\uu{\Sp}{*}{\VPLEN}{}$ and $\Sq\VQLEN\uu{\Sq}{*}{\VQLEN}{}$, respectively \citep{bib:waleffe1992nature}.
The energy flux between the three triad legs is fixed for a given triad and is determined by the terms $(\Sp\VPLEN - \Sq\VQLEN), (\Sq\VQLEN - \Sk\VKLEN)$
and $(\Sk\VKLEN - \Sp\VPLEN)$ in  \eqnref{eqn:triadeqns}, while the average flux direction (to or from a leg) is determined by the
sign of the three-wave correlator $\langle \uu{\Sk}{*}{\VKLEN}{}\uu{\Sp}{*}{\VPLEN}{}\uu{\Sq}{*}{\VQLEN}{}\rangle+c.c.$.

\citet{bib:waleffe1992nature} suggested that a linear instability analysis would predict the average energy flux directions within helical triad interactions by assuming that energy, on average, flows out of the most unstable wave mode and into the other two. Evaluating the stability of the fixed points $\lbrace \uf{\Sk}{\VKLEN}, \uf{\Sp}{\VPLEN},\uf{\Sq}{\VQLEN}\rbrace = \lbrace U_0,0,0 \rbrace,\lbrace 0,U_0,0 \rbrace,\lbrace 0,0,U_0 \rbrace$ using \eqnref{eqn:triadeqns}, one finds the unstable mode is determined by $(\Sq\VQLEN - \Sk\VKLEN)(\Sk\VKLEN - \Sp\VPLEN)>0$.
This criterion implies the smallest leg is unstable in G1 and G3 interactions, suggesting that these interactions contribute with a forward energy cascade (F-class interactions), while for G2 and G4 
the middle leg is unstable, suggesting part of the energy flux is reversed. 
In G4, only same-signed helical modes interact, implying both positive and negative helicities, $H^+ = \sum_\vect{k} \VKLEN\vert u_{+}(\VK) \vert^2$ and $H^- = \sum_\vect{k} \VKLEN\vert u_{-}(\VK) \vert^2)$, are separately conserved. 
As such, G4 interactions can be regarded analogous to enstrophy-conserving 2D interactions, and, consequently, should contribute with a reversed energy cascade (R-class interactions). This was recently indeed found to be the case numerically \citep{bib:biferale2012inverse}.  Note that the 2D analogy argument for why G4 interactions should exhibit a reversed energy cascade is different from that of the instability assumption. 
Lastly, in G2 interactions, positive and negative helicity components do interact, thus breaking the helicity-enstrophy analogy for explaining the mixed F- and R-class nature of G2.  

Here we argue that the mixed F- and R-class nature of G2 interactions is determined by a quantity different from energy and helicity, which too is
conserved within a single triad interaction \eqnref{eqn:triadeqns}, but depends on triad shape. 
This new "pseudo-invariant" is thus unlike energy and helicity not a globally conserved quantity (across all triad interactions) because of its shape dependency.
Our conjecture, then, is that the energy cascade, within subsets of identically shaped triads, should transition from forward (F-class) to reverse (R-class) depending on whether energy or the pseudo-invariant is dominant at the dissipation scale.
To realise this, consider the spectral pseudo-invariant quantity defined as
\gdef\expoalpha{       \alpha(\VKLEN,\VPLEN,\VQLEN)}
\gdef\expoalphageneral{\alpha(\VKLEN,\VPLEN,\VQLEN)}
\gdef\expoalphashort{\alpha}
\gdef\angk{\theta}
\gdef\angp{\theta\sprime}
\gdef\angq{\theta\dprime}
\gdef\NSEeps#1#2#3#4{\epsilon_{#1}^{#2,#3,#4}}
\gdef\NSEepsdef#1#2#3#4#5{(#3-#5\lambda^#2)/(\lambda^#1-\Sp\Sq\lambda^#2)}
\gdef\Bsymb{E^{(\alpha)}}
\gdef\Bdef#1#2#3#4{    #4^\genexpEsymb\left(\vert u_{#1}(#3) \vert ^2 + \vert u_{#2}(#3) \vert^2\right)}
\gdef\Bdefhalf#1#2#3#4{#4^\genexpEsymb \vert u_{#1}(#3) \vert ^2}
\gdef\expocond{\genexpEsymb\in \mathbb{R}\setminus 0}
\gdef\expocondgeneral{\genexpEsymb\in \mathbb{R}}
\begin{align}
\Bsymb(\VK)= \Bdef{+}{-}{\VK}{\VKLEN} ,\;\;\alpha \in  \mathbb{R},
\end{align}
which is analogous to the spectral energy density $E(\VK)=|u_+(\VK)|^2+|u_-(\VK)|^2$.   
This quantity is conserved by triad interactions governed by \eqnref{eqn:triadeqns} if $d_t\left(\Bsymb(\VK)+\Bsymb(\VP) +\Bsymb(\VQ)\right) =0$, implying
\gdef\VPVKRATIO{\frac{\VPLEN}{\VKLEN}}
\gdef\VQVKRATIO{\frac{\VQLEN}{\VKLEN}}
\begin{align}
\left(\Sp\VPVKRATIO - \Sq\VQVKRATIO\right)  + \left(\VPVKRATIO\right)^\genexpEsymb\left(\Sq\VQVKRATIO - \Sk\right)  +  \left(\VQVKRATIO\right)^\genexpEsymb\left(\Sk - \Sp\VPVKRATIO\right)  = 0, \label{eqn:Bcond_long}
\end{align}
which is trivially fulfilled for any triad when $\alpha=0$ (i.e. energy).
As a function of triad shape, given by the relative leg sizes $\VPLEN/\VKLEN$ and $\VQLEN/\VKLEN$, the left-hand side of \eqnref{eqn:Bcond_long} consists of a constant term and two monotonically increasing/decreasing terms. 
The existence of a non-trivial real solution, $\alpha \ne 0$, for a given triad shape $\lbrace\VKLEN,\VPLEN,\VQLEN\rbrace$ and interaction group $\sigmastate$ therefore requires the signs of the coefficients of the two last terms in \eqnref{eqn:Bcond_long} be opposite. 
Note that no more than \textit{one} non-trivial real solution can exist. 
It follows that only G2 and G4 interactions can have non-trivial solutions to \eqnref{eqn:Bcond_long}.
For G4, $\genexpEsymb=1$ is the solution for any triad, corresponding to the global inviscid conservation of helicity, as expected. 
For G2, the solution $\expoalphashort = \expoalpha$ is triad shape dependent.
\begin{figure}[t]
\includegraphics[scale=0.83]{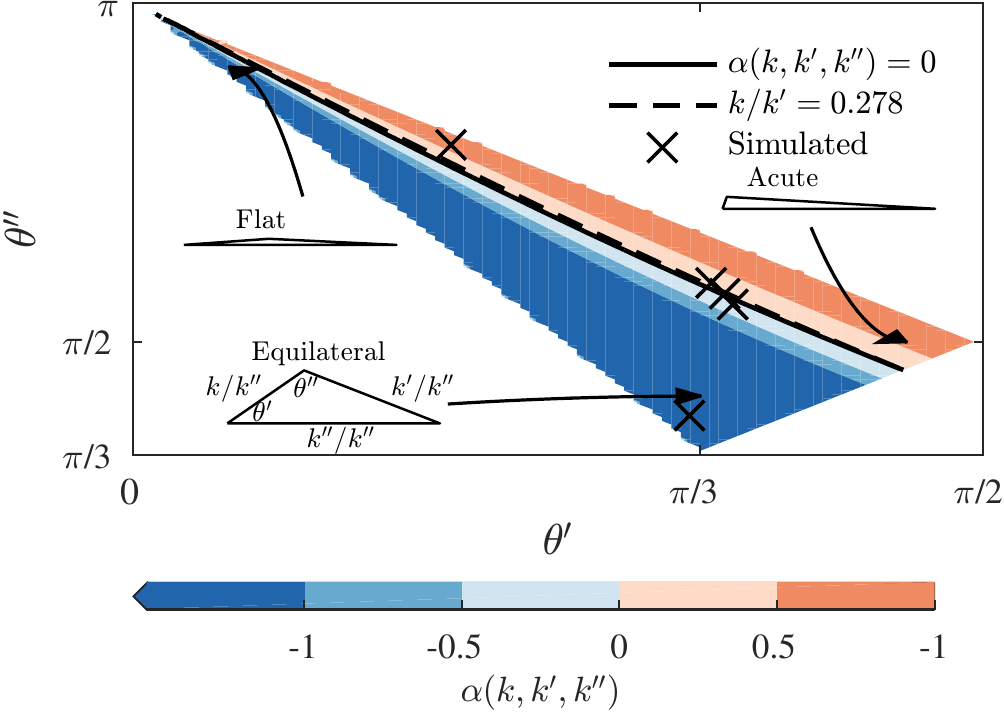} 
\caption{
G2 $\expoalphashort$-solutions as a function of triad shape given by the two interior angles $\angp$ and $\angq$. Overlayed are the contours for $\expoalphashort = 0$ using equation \eqnref{eqn:transitioncond} (full line) and $\VKLEN/\VPLEN=0.278$ (dashed line) together with the specific triad geometries simulated in this study (crosses). \label{fig:m2prediction}
\vspace*{-1em}
 }
\end{figure}
FIG. \ref{fig:m2prediction} shows the numerically solved G2-solutions for all possible triad geometries by expressing each triad in terms of the two interior angles $\angp$ and $\angq$ using the Sine rule $\VPLEN/\VKLEN = \sin(\angp)/\sin(\pi-\angp-\angq)$ and $\VQLEN/\VKLEN=\sin(\angq)/\sin(\pi-\angp-\angq)$. By expressing triads in terms of interior angles, all non-congruent triads are mapped onto the coloured triangle in FIG. \ref{fig:m2prediction}. 
The region near the top left corner corresponds to flat triangles ($\VKLEN\approx \VPLEN\approx \VQLEN/2$) whereas the region near the rightmost corner corresponds to acute triangles ($\VKLEN\ll \VPLEN\approx \VQLEN$), the latter representing the space of non-local triad interactions.  Finally, the lower corner corresponds to the equilateral triangle ($\VKLEN=\VPLEN=\VQLEN$), representing local triad interactions.  

For G2 triad shapes fulfilling
\begin{align}
\log{(\VQLEN/\VKLEN)}/(1+\VQLEN/\VKLEN) = \log{(\VPLEN/\VKLEN)}/(1+\VPLEN/\VKLEN)
\label{eqn:transitioncond}
\end{align}
(taking $d/d\alpha\vert_{\alpha=0}$ of equation \eqnref{eqn:Bcond_long}), the trivial and non-trivial solutions collapse to the single solution $\alpha=0$. 
Because the ratio of the spectral pseudo-invariant density to energy scales as $k^\alpha$ (growing with $k$ for $\alpha>0$), the subset of G2 triad interactions having $\alpha>0$ (red in FIG. \ref{fig:m2prediction}) may be regarded analogous to enstrophy-conserving interactions in 2D turbulence. 
Note that these triads interactions correspond to non-local interactions. 

In \citet{bib:waleffe1992nature} the direction of the spectral energy flux was calculated for the four helically decomposed triad interaction types (G1-G4) based on the Kolmogorov similarity assumption. A change of sign in the calculation of the G2 energy flux direction suggested that G2 triads with leg ratios of $\VKLEN/\VPLEN<0.278$ should contribute to a reverse energy cascade. This constraint is marked by the dashed line in FIG. \ref{fig:m2prediction}. Note that the constraint proposed in \citet{bib:waleffe1992nature} (dashed line) is close to our proposed constraint $\alpha=0$ (full line).

In order to test our conjecture, we apply our newly constructed shell model \citep{bib:rathmannditlevsen2016} (source available at \url{https://github.com/nicholasmr/rdshellmodel}). The model is somewhat similar to the helical 
Sabra model \cite{bib:DePietro2015}, but additionally allows coupling the four interactions groups (G1-G4) and multiple triad shapes through coupling weights derived directly from \eqnref{eqn:Helical_NSE}. 
Using this model, it is straight forward to perform "spectral surgery" as proposed by \citep{bib:biferale2012inverse,bib:biferale2013split,bib:sahoo2015,bib:DePietro2015} in order to investigate the (isolated) behaviour of G2 interactions. 
Considering only fixed-shaped G2 interactions the shell model takes the form
\gdef\expoalpha{\alpha(k_n,k_{n+p},k_{n+q})}
\gdef\expoalphashort{\alpha}
\gdef\expohatalpha{\hat{\alpha}(k_n,k_{n+p},k_{n+q})}
\gdef\expohatalphashort{\hat{\alpha}}
\gdef\forcingshellnr{108}
\gdef\LLgenexpEpseudo{\genexpEsymb}
\gdef\LLgenexpHpseudo{\genexpHsymb}
\begin{align}
&(d_t+\nu k_n^2 +\nu_Lk_n^{-2}) u_s (k_n)= f_s(k_n) + s k_n[u_{-s}^*(k_{n+p} )u_{-s}(k_{n+q} ) + \nonumber\\
&\frac{\epsilon_{p,q}}{\lambda^p} u_{-s}^*(k_{n-p} )u_{s}^*(k_{n+q-p}) + \frac{\epsilon_{p,q}+1}{\lambda^q} u_{-s}(k_{n-q} )u_{s}(k_{n-q+p} )],\label{eqn:NSSM_compact_deriv}
\end{align}
where $\epsilon_{p,q} = (1+\lambda^q)/(\lambda^p-\lambda^q)$, $f_s(k_n)$ is the forcing at wave number $k_n$, and the linear terms $\nu k_n^2 u_s (k_n)$ and $\nu_Lk_n^{-2}u_s (k_n)$ are viscous dissipation and
a drag term, respectively, the latter added in the usual way to remove energy at large scales. 
The scalars $k_n=k_0\lambda^n$, where $n=0,\cdots, N$, represent the exponentially distributed shell wave numbers resolved, $\pqset\in \mathbb{N}$, $k_0\in\mathbb{R}_+$ and $\lambda\in\;]1,(1+\sqrt{5})/2] = \;]1,\varphi]$. The golden ratio $\varphi$ is the upper limit such that any set of nearest neighbour waves fulfil the triangle inequality as required by the NSE. 

The integers $\pqset$ can be related to any triangular shape through the Sine rule.  
The possible resolved triad shapes depend therefore on the combination of $\lbrace \lambda, p, q\rbrace$:
For $\lambda\rightarrow 1$ any triad geometry may be constructed for large/small enough values of $\lbrace p,q\rbrace$, while for $\lbrace \lambda,p,q\rbrace = \lbrace \varphi,1,2\rbrace$ triads collapse to a line. 
Thus, for each chosen set of $\lbrace \lambda,p,q\rbrace$ the shell model consists independently of scale $k_n$ only of fixed-shaped triad interactions.

\gdef\KIDXs{n}
The non-linear terms in \eqnref{eqn:NSSM_compact_deriv} conserve both energy $\Einvarreal = \sum_{n=0}^N  \modalcontr{+}$ and helicity $\Hinvarreal = \sum_{n=0}^N  k_n \modalcontr{-}$.
Each $\pqset$-configuration of the model ($\lambda$ hereafter assumed fixed)
additionally conserves the pseudo-invariants $\Bsymb = \sum_{n=0}^N  k_n^\genexpEsymb\modalcontr{+}$
in complete analogy to \eqnref{eqn:Bcond_long} for the NSE. 

\gdef\Ecorr#1{\Delta_{#1,p,q}^{}}
\gdef\pseudoEcorr#1{\Delta_{#1,p,q}^{\Einvarpseudo}}

\gdef\KIDXs{m}
The non-linear spectral energy flux through the $n$-th shell is for a given $\pqset$-configuration of the model given by \citep{bib:rathmannditlevsen2016}
\gdef\KIDXs{n}
$
\Pi(k_n) = \sum_{{m=n+1}}^{{n+q}} \Ecorr{m} + \Wepssubmodel\sum_{{m=n+1}}^{{n+q-p}} \Ecorr{m} 
$
where
\gdef\KIDXs{m}
$\Ecorr{\KIDXs} =
2 k_{\KIDXs-q}\Re[
  u^{*}_{+}(k_{\KIDXs-q})u^{*}_{-}(k_{\KIDXs-q+p})u^{}_{-}(k_{\KIDXs})  
- u^{*}_{-}(k_{\KIDXs-q})u^{*}_{+}(k_{\KIDXs-q+p})u^{}_{+}(k_{\KIDXs})]
.
$

Simulations were conducted with $\lambda = 1.1$, $k_0=1$, $N=223$, $\nu = \SI{1e-12}{}$ and $\nu_L = \SI{1e2}{}$. 
Five different sets of $\pqset$ were chosen: $p=\lbrace 1, 12, 13, 14, 22\rbrace$ with $q=p+1$, corresponding to $\alpha=\lbrace -30.94, -0.15, 0.01, 0.15, 0.69\rbrace$ (crosses in FIG. \ref{fig:m2prediction}). 
In all simulations the forcing 
$f^\pm_{n_{\text{f}}} = (1+i)/\ush{n_{\text{f}}}{\pm}{1}$
was applied to both helical components at shell $n_{\text{f}} = \forcingshellnr$, supplying a constant energy input of $\meanenergy = 4$.

\begin{figure}[b]
\centering \includegraphics[scale=0.63]{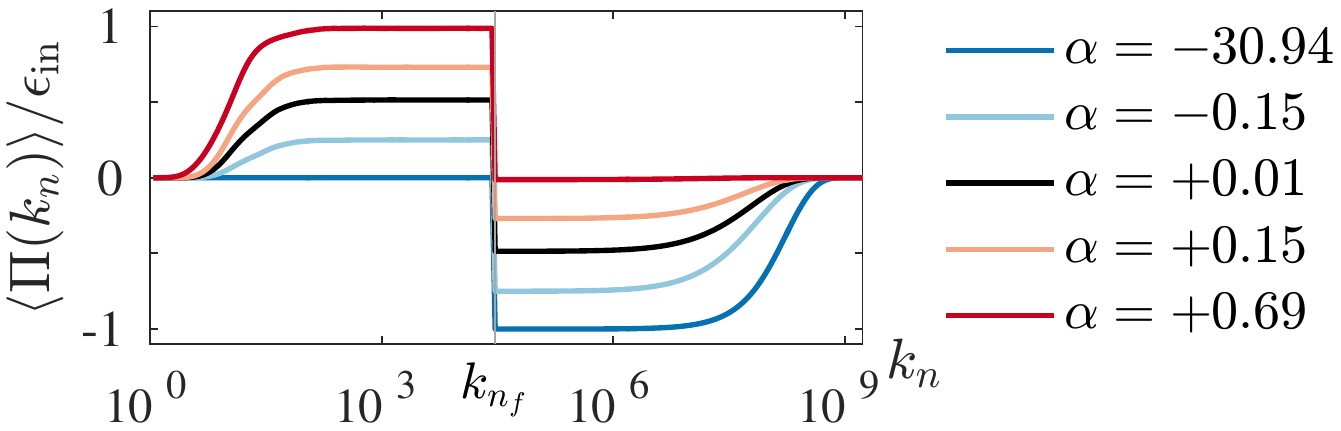} 
\caption{Simulated G2 spectral energy fluxes of the triad geometries ($\alpha$-values) considered.
\label{fig:mains_results}}
\end{figure}

FIG. \ref{fig:mains_results} shows the simulated spectral energy fluxes. 
The blue curves show the resulting energy fluxes for the model configured with triad shapes having $\alpha<0$, in which case energy should exhibit a forward cascade. The red curves show the opposite with $\alpha>0$, namely a 2D-like reversed energy cascade and a forward cascade of the enstrophy-like pseudo-invariant (latter not shown). As the cascade directions for the energy and the pseudo-invariant interchange at
$\alpha=0$, we expect a split forward/reversed energy cascade to develop, which is indeed found to be the case (black curve in FIG. \ref{fig:mains_results}).  

The average energy spectra are found to accommodate the transfer directions by scaling K41-like as $\sim k^{-2/3}$ wherever the energy cascade dominates, i.e. $k_n > k_{n_{\text{f}}}$ for local triads and $k_n < k_{n_{\text{f}}}$ for non-local triads (not shown).

\gdef\GWEIGHT#1{I_{#1}} 
\gdef\QWEIGHT{Q} 
\gdef\THETAP{\theta\sprime}
\gdef\THETAQ{\theta\dprime}
\gdef\VPVQjacobian{k^2\sin(\THETAP)\sin(\THETAQ){\left(1+\cos(\THETAP+\THETAQ\right)^2)/}{\sin(\THETAP+\THETAQ)^4}}
\gdef\VPVQjacobiandet{\vert \det J \vert}
\gdef\dVPdVQ{d\VPLEN d\VQLEN}
\gdef\dTHETAPdTHETAQ{d\THETAP d\THETAQ}
\begin{figure}[t]
\includegraphics[scale=0.83]{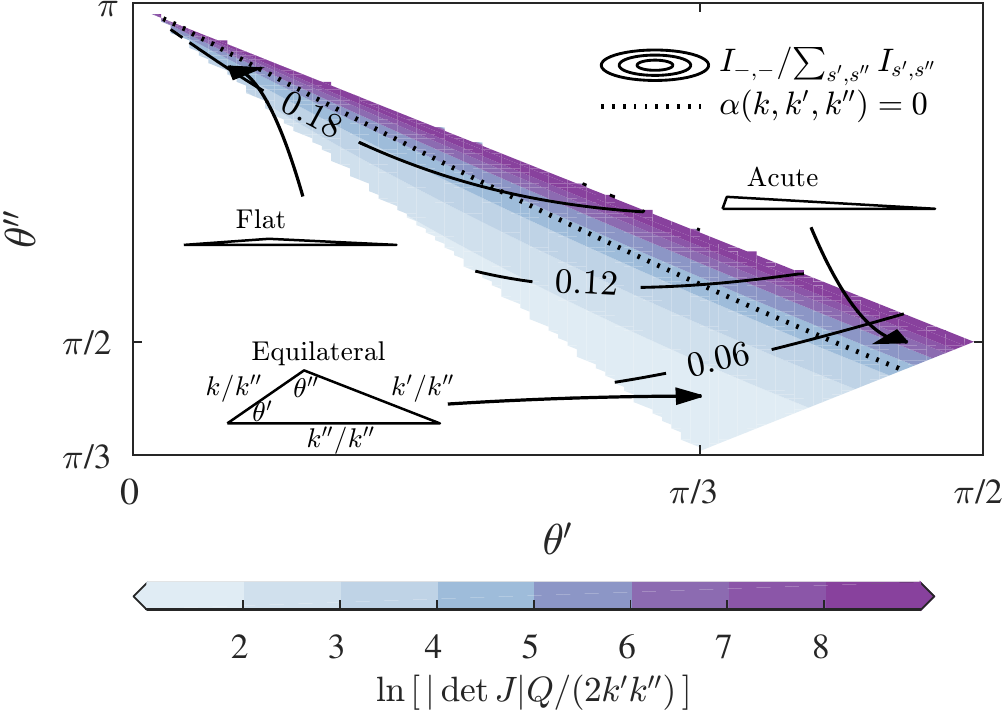} 
\caption{Density of triads multiplied by their coupling coefficient (coloured contours) compared to the G2 coupling coefficient normalize against the G1-G4 coupling coefficients (black contours). \label{fig:triaddensity}
\vspace*{-1em}
 }
\end{figure}

The importance of the "hidden" reverse energy cascade carried by G2 R-class interactions ($\alpha>0$), which are mostly non-local, depends (i) on the number of G2 R-class triads compared to the number of G2 F-class triads, and (ii) the magnitudes of the G2 R-class coupling coefficients in \eqnref{eqn:Helical_NSE}  compared to the coefficients for G1, G3 and G4.
To estimate (i), consider the continuous version of \eqnref{eqn:Helical_NSE} where the triad sum becomes an integral over $\dVPdVQ$.
In terms of $\THETAP$ and $\THETAQ$, the corresponding density of triads within the element $\dVPdVQ$ is given by the transformation $\dVPdVQ = \VPVQjacobiandet\, \dTHETAPdTHETAQ$, where $J=\partial \Phi$ is the Jacobiant of the transformation  $k'=\Phi'(\theta',\theta'')$ and $\VPVQjacobiandet = \VPVQjacobian$. Thus, the number of G2 R-class triads far exceeds the number of F-class triads in the limit of large inertial ranges ($\mathrm{Re}\rightarrow \infty$) since the acute triad limit $\VPLEN,\VQLEN\rightarrow \infty$ implies $\sin(\THETAP+\THETAQ)\rightarrow 0$ and therefore a large density of non-local triads.
To estimate (ii), consider the relative (normalized) magnitudes of the G2 coupling coefficients given by $\GWEIGHT{-,-}/\sum_{\Sp,\Sq}\GWEIGHT{\Sp,\Sq}$ where $\GWEIGHT{\Sp,\Sq} = \vert (\Sk\VKLEN+\Sp\VPLEN+\Sq\VQLEN)(\Sp\VPLEN-\Sq\VQLEN) \vert$, which originates from \eqnref{eqn:Helical_NSE} since $\vert (\Sp\VPLEN - \Sq \VQLEN)\,g \vert = \GWEIGHT{\Sp,\Sq} {\QWEIGHT}/{(2\VKLEN\VPLEN\VQLEN)}$ where $\QWEIGHT = (2\VKLEN^2\VPLEN^2 + 2\VPLEN^2\VQLEN^2 + 2\VQLEN^2\VKLEN^2 -\VKLEN^4 - \VPLEN^4 - \VQLEN^4)^{1/2}$ \citep{bib:waleffe1992nature}.
Assuming $k=1$ without loss of generality, the coloured contours in FIG. \ref{fig:triaddensity} show the triad density, $\VPVQjacobiandet$, multiplied by the part of the coupling coefficient common between G1-G4, ${\QWEIGHT}/{(2\VPLEN\VQLEN)}$, indicating the density of triads increases faster with non-locality than the decrease in coupling strength. This suggests non-local interactions become increasingly important to the extent that the inertial range is long enough for them to be resolved. Overlayed FIG. \ref{fig:triaddensity} is the relative G2 coupling magnitudes (black lines), suggesting G2 R-class interactions should, overall, play an important role in the helically decomposed dynamics of flat/semi-acute triads.  
 

In conclusion, we presented an alternative classification to linear triad stability analysis \citep{bib:waleffe1992nature} for explaining the nature of the eight elementary non-linear interactions of the spectral Navier--Stokes equation in the helical basis. By showing a subset of interactions conserve new enstrophy-like blocking quantities depending on triad geometry, the apparent complicated nature of the second group (G2) of helical interactions (FIG. \ref{fig:helicalcouplings}) may be explained in terms of physically conserved quantities analogous to enstrophy in 2D turbulence. 

\bibliographystyle{apsrev4-1}
\bibliography{mclass_blocking}

\end{document}